\documentclass{jjap3}
\usepackage{bm,color}
\usepackage{here}

\title{Memory functions of magnetic skyrmions}
\author{W. Koshibae$^{1}$, Y. Kaneko$^{1}$, J. Iwasaki$^{2}$, M. Kawasaki$^{1,2}$, 
Y. Tokura$^{1,2}$ and N. Nagaosa$^{1,2}$}
\inst{$^1$ RIKEN Center for Emergent Matter Science (CEMS), Wako, Saitama 351-0198, Japan
\\
$^2$ Department of Applied Physics and Quantum-Phase Electronics Center (QPEC), 
The University of Tokyo, 7-3-1, 
Hongo, Bunkyo-ku, Tokyo 113-8656, Japan
}
\abst{Magnetic skyrmion, a swirling spin texture, 
in chiral magnets 
is characterized by (i) nano-scale size ($\sim$1nm -- 100nm), 
(ii) topological stability, and 
(iii) gyro-dynamics. 
These features are shown to be advantageous for 
(a) high-density data-storage, 
(b) nonvolatile memory, and (c) ultra-low current and energy cost manipulation,
respectively. By the numerical simulations of Landau-Lifshitz-Gilbert equation, 
the elementary functions of skyrmions are demonstrated aiming 
at the design principles of skyrmionic memory devices.  
}

\begin{document}
\maketitle

\section{Introduction}
\begin{figure}[t]
\begin{center}
\includegraphics[scale=0.6,angle=0,clip]{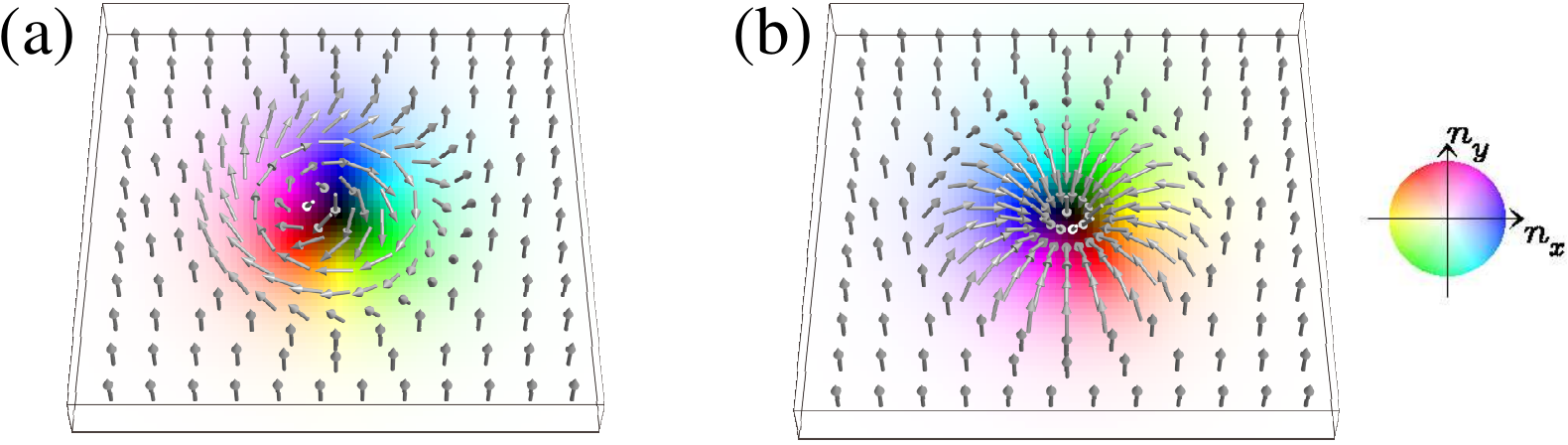}
\end{center}
\caption{(Color online) Schematic illustration of a skyrmion.  
(a) Bloch skyrmion.  From the center to the rim, the rotating magnetic moments are 
always perpendicular to the radial vector.  
(b) N\'eel skyrmion.  The rotating plane of magnetic moments is  
perpendicular to that of Bloch skyrmion. 
The color map (right) specifies the in-plane components of the magnetic moment 
and the brightness of the color represents the out-of plane component. 
}
\label{fig1}
\end{figure}

High density, high speed, low-energy cost nonvolatile memory devices are regarded as 
an essential element of the next-generation electronics. 
Many candidates have been explored up to now in this respect, such as 
Magnetoresistive Random Access Memory (MRAM), 
Phase change Random Access Memory (PRAM), 
Resistance Random Access Memory (ReRAM), 
etc.~\cite{Fujisaki1,Fujisaki2,Zahurak,Ikegami,Boniardi,Lam}  
In addition to these existing memory devices, 
we propose in this paper the memory function utilizing skyrmions (Sks)~\cite{reviewNT}  
as the new promising candidate which has many superior characteristics. 
Figure \ref{fig1} shows schematically the magnetic configuration of an Sk in magnet. 
It is a vortex-like structure in the ferromagnetic (F) background with the magnetic moments 
pointing down at the core and swirling in the intermediate region. 
Because the winding plane is perpendicular to the radial direction in Fig.~\ref{fig1}(a), 
it is called Bloch Sk.  
It is noted that an Sk has a finite size $\xi$ typically of the order of 1--100nm, 
in contrast to a vortex or a meron,~\cite{Ezawa,YanZhou} which is 
``half" of an Sk with the swirling in-plane components of the magnetic moments extending to infinity.  
This size $\xi$ is determined by the ratio of the relativistic 
Dzyaloshinskii-Moriya (DM) interaction~\cite{Dzyaloshinskii,Moriya} $D$ 
and the exchange coupling $J$ as $\xi\sim(J/D)a$ with $a$ being the lattice constant. 
The DM interaction is allowed in non-centrosymmetric chiral magnets such as 
MnSi~\cite{Muhlbauer09,Ishikawa1,Ishikawa2,Lebech1,Pfleiderer,Janoschek}, 
Fe$_{1-x}$Co$_x$Si~\cite{Munzer10,YuXZ10N,Ishimoto,Grigoriev1,Grigoriev2,Onose}, 
FeGe~\cite{Lebech2,Uchida,YuXZ11,Wilhelm}, 
and Mn$_{1-x}$Fe$_x$Ge~\cite{Shibata} and Cu$_2$OSeO$_3$~\cite{Seki12}. 
Intuitively speaking, DM interaction prefers the twisted spin configurations 
while the external magnetic field stabilizes the F one. 
To reconcile these two interactions, the spiral magnetic configurations are 
preferred and as the special form among them, the skyrmion crystal (SkX) 
is realized in the intermediate region of the external magnetic field. 
In the multilayered magnet, 
the uniform DM vector is anticipated to exist along the interface of the layers, and hence 
a N\'eel Sk shows up~\cite{Heinze} 
which is schematically shown in Fig.~\ref{fig1}(b).  
These N\'eel and Bloch Sks are in the same topology class as discussed below.  

The topology of Sk is characterized by an integer $N_{sk}$ 
called skyrmion number which counts how many times the direction of 
the magnetic moments $\bm n_{\bm r}$ 
wrap the unit sphere.  Both Bloch and N\'eel Sks have the same $N_{sk}$.  
This topological integer does not change as long as 
$\bm n_{\bm r}$ remains slowly varying compared with the lattice constant $a$, 
i.e., $\bm n_{\bm r}$ is viewed as a continuous function of the spatial coordinates $\bm r$. 
This provides the stability of an Sk as an emergent particle composed of many magnetic moments. 
For example, an Sk created in the F background has 
a very long lifetime even though it is a meta-stable state. 
In other words, the discontinuity of the magnetic configuration must be 
introduced to create or annihilate Sks, which we call ``topological surgery". 

As for the dynamical properties of the skyrmionic systems, 
it is noted that the effective magnetic field is associated with the Sk. 
Namely, the solid angle subtended by the spins leads to the Aharanov-Bohm effect 
and acts as an emergent magnetic field both for the Sk itself 
and conduction electrons coupled to the Sk~\cite{reviewNT}. 
The equation of motion under the spin polarized electron current 
$\bm j_s$ (of which sign is opposite to the electric current $\bm j$) 
for the center-of-mass motion of an Sk reads 
\begin{equation}
m\frac{d\bm v_d}{dt}+\bm G\times\left(\bm j_s-\bm v_d\right)
+\kappa\left(\alpha\bm v_d-\beta\bm j_s\right)
=-\nabla U.
\label{eqn:1}
\end{equation}
Here $(X,Y)$ is the centre of mass coordinate and 
$\bm v_d=(\dot{X},\dot{Y})$ is its velocity.    
The mass $m$ of the Sk originates from the deformation of the moving Sk 
compared with the static solution.  
Recently, this mass $m$ has been studied extensively, 
and it is found that $m$ is almost zero 
when the motion is driven by the electric current~\cite{Schuette}. 
The gyrovector $\bm G=(4\pi N_{sk})\bm e_z$ ($\bm e_z$: unit vector along $z$-direction) 
describes the gyro-dynamics of the Sk analogous 
to that of a charged particle under magnetic field. 
A dimensionless constant $\kappa$ is of the order of unity, 
$\alpha$ denotes the Gilbert damping constant, 
$\beta$ represents the coefficient of the non-adiabatic effect, 
$U$ is the potential and $-\nabla U$ is the force acting on the Sk, 
e.g., those from the boundary, gradient of the magnetic field, and the impurities. 
As for the conduction electrons, they are also subject to 
the effective magnetic flux corresponding to $\phi_0=2\pi\hbar/e$ per one Sk 
in the limit of strong spin-charge coupling~\cite{reviewNT}.

In this paper, we explore the memory functions of Sks utilizing 
these physical properties by solving the Landau-Lifshitz-Gilbert (LLG) 
equation numerically. In section \ref{Sec:section2}, 
several elementary functions of Sks are revealed, 
and some models of memory devices are proposed in section \ref{Sec:somemodels}. 
Section \ref{Sec:summary} is devoted to the summary and perspectives for future Sk memory.

\section{Elementary functions for Sk memory}
\label{Sec:section2}

The model Hamiltonian for the chiral magnets defined on the 
two-dimensional square lattice is given by  
\begin{align}
\mathcal{H}=&-J\sum_{\bm r} {\bm n}_{\bm r} \cdot 
\left( {\bm n}_{{\bm r}+{\bm e_x}}+{\bm n}_{{\bm r}+{\bm e_y}} \right) 
\notag \\
&+\sum_{\bm r} \left( 
 {\bm n}_{\bm r} \times {\bm n}_{{\bm r}+{\bm e_x}} \cdot {\bm D_{1,\bm r}}
+{\bm n}_{\bm r} \times {\bm n}_{{\bm r}+{\bm e_y}} \cdot {\bm D_{2,\bm r}} \right) 
\notag \\
&-\sum_{\bm r} h_{\bm r} n_{z,\bm r}. 
\label{eqn:Hamiltonian}
\end{align}
Here we take the lattice constant $a$ as the unit of length.  
The DM vectors $\bm D_1$ and $\bm D_2$ specify 
the helicity (direction of vortex) and type (Bloch or N\'eel) of Sk, i.e., 
$\bm D_{1,\bm r}=D\bm e_x$ and $\bm D_{2,\bm r}=D\bm e_y$ for Bloch Sk 
and 
$\bm D_{1,\bm r}=-D\bm e_y$ and $\bm D_{2,\bm r}=D\bm e_x$ for N\'eel Sk,  
where $\bm e_x$ and $\bm e_y$ 
are the unit vectors in the $x$- and $y$- directions, respectively,  
and the sign of $D$ is responsible for the helicity.  
The combination of the exchange interaction $J$ and $D$
produces the single-$q$ helical state with 
$q = \tan^{-1}(D/J)$ under zero magnetic field.
Usually $D\ll J$, and hence the helix period 
$\frac{2 \pi}{q} \cong \frac{2 \pi J}{D}$ 
is much longer than the lattice constant,
and hence the continuum approximation is justified.  
The critical magnetic field $h_c$~\cite{Han1,Mochizuki,Iwasaki1} separating the
SkX state and the F state in the ground state
is $h_c \approx 0.78\times(D^2/J)$.  
In the numerical study shown below, we apply a bias magnetic field $h_0$ being 
slightly larger than $h_c$ where the perfect F state 
is the ground state and therefore an Sk is defined as a meta-stable object.

We study the creation, annihilation and drive of 
an Sk for various parameters in the Hamiltonian 
Eq.~(\ref{eqn:Hamiltonian}), system size $L$ and boundary conditions, 
using the 
stochastic Landau-Lifshitz-Gilbert equation given by 
\begin{equation}
\frac{{\rm d} {\bm n}_{\bm r}}{{\rm d} t}=
\gamma\left[- \frac{\partial \mathcal{H}}{\partial {\bm n}_{\bm r}}
+{\bm h}_{T(\bm{r})}(t)\right]
\times {\bm n}_{\bm r}
+\alpha {\bm n}_{\bm r} \times
\frac{{\rm d}{\bm n}_{\bm r}}{{\rm d} t}
-\left(\bm j_{\bm r}\cdot\nabla\right)\bm n_{\bm r}
+\beta\left[\bm n_{\bm r}\times\left(\bm j_{\bm r}\cdot\nabla\right)\bm n_{\bm r}\right].
\label{eqn:LLG}
\end{equation}
Here the Gaussian noise field 
${\bm h}_{T(\bm{r})}(t)=(h_{T(\bm{r}),x}(t),h_{T(\bm{r}),y}(t),h_{T(\bm{r}),z}(t))$
with the statistical properties $\langle h_{T(\bm{r}),\nu}(t)\rangle =0$ and 
$\langle h_{T(\bm{r}),\mu}(t)h_{T(\bm{r}'),\nu}(t')\rangle
=2k_BT(\bm{r})\alpha\delta_{\mu\nu}\delta(\bm{r}-\bm{r}')\delta(t-t')$  
($\mu,\nu=x,y,z$) is introduced to examine  
the effect of local heating by a spatially dependent 
temperature $T(\bm{r})$ on the Langevin dynamics of the system.  
We use the Heun scheme to solve
this equation~\cite{GPalacios98,MersenneTwister}.
The last two terms represent the spin-transfer-torque (STT) 
induced by the (spin-polarized) electric current density $\bm j_{\bm r}$.  
Below, the time and $j=|\bm j_{\bm r}|$ are 
measured in units of $1/(\gamma J)$ and 
$2e\gamma J/pa^2$ 
($p$: polarization of magnet), 
and the units are typically $1/(\gamma J) \sim 0.7$ ps and 
$2e\gamma J/pa^2\sim1.0\times 10^{13}$ A/m$^2$, respectively,   
if we assume $\gamma=g_s\mu_B/\hbar$ ($g_s$: electron spin $g$-factor, $\mu_B$: Bohr magneton), 
$J \sim 10^{-3}$ eV, $p=0.2$ and $a=5$\AA. 

\vskip1cm

\subsection{Write and Erase methods}
\label{write/erase}

\subsubsection{Heat-control Sk memory.}
The Sk creation/annihilation in the F background is 
a topological transition of the magnetic texture 
associated with the singular configurations of magnetic moments.  
These singular configurations can be realized  
by local heating~\cite{koshibae} (see Fig.~\ref{heat} and the Movie 1 in the Supplementary Informations).  
Figure \ref{heat}(a) is a closeup (100$\times$100 in $L=300\times300$)  
of the F state where all magnetic moments are perpendicular to the plane.  
A parameter set $\{J=1.0, D=0.15, h_{\bm r}=h_0=0.02, \alpha=0.01\}$ and periodic 
boundary condition (PBC) are used.  
The Sk creation is achieved by applying 
a local heat corresponding to the temperature $k_BT=1.5J$ 
($\sim T_c$) with 
a duration $t_{heating}=200$ inside the red circle ($15$ in radius) in Fig.~\ref{heat}(a), 
and Fig.~\ref{heat}(b) shows the created Sk. 
This heating effect is introduced in Eq.~(\ref{eqn:LLG}) by the Gaussian noise field 
${\bm h}_{T(\bm{r})}(t)$, 
i.e., $k_BT(\bm{r})(t)=1.5J$ in the red circle in Fig.~\ref{heat}(a) for $0\leq t\leq t_{heating}$ 
and $k_BT(\bm{r})(t)=0$ otherwise.  
By local heating, the magnetic moments at around the heat-spot are strongly 
disordered and constitute a nucleation center for the Sk creation.  
This condition depends on the heat intensity, the heat-spot size and also the 
Gilbert damping constant $\alpha$.  For smaller $\alpha$, the dynamics of magnetic moments becomes 
more active.  The active dynamics, however, is rather destructive for Sk, i.e., 
too much heat is $not$ suitable for Sk writing.  
Conversely, the strong heating is useful to erase the Sk.  
Figures ~\ref{heat}(c)$\rightarrow$(d) demonstrate the Sk erasing 
by the stronger heating (see also the Movie 2 in the Supplementary Informations):  
An Sk is prepared at the initial state as shown in Fig.~\ref{heat}(c).  
The heat intensity $k_BT=1.5J$ and heating duration $t_{heating}=200$ 
are the same with those used for the Sk writing in Fig~\ref{heat}(a)$\rightarrow$(b), but 
the larger heat-spot with radius 20 is used  
(the size is shown by the blue circle in Fig.~\ref{heat}(a)).  
This heat destroys the Sk as shown in Fig.~\ref{heat}(d).  

\begin{figure}[t]
\begin{center}
\includegraphics[scale=0.4,angle=0,clip]{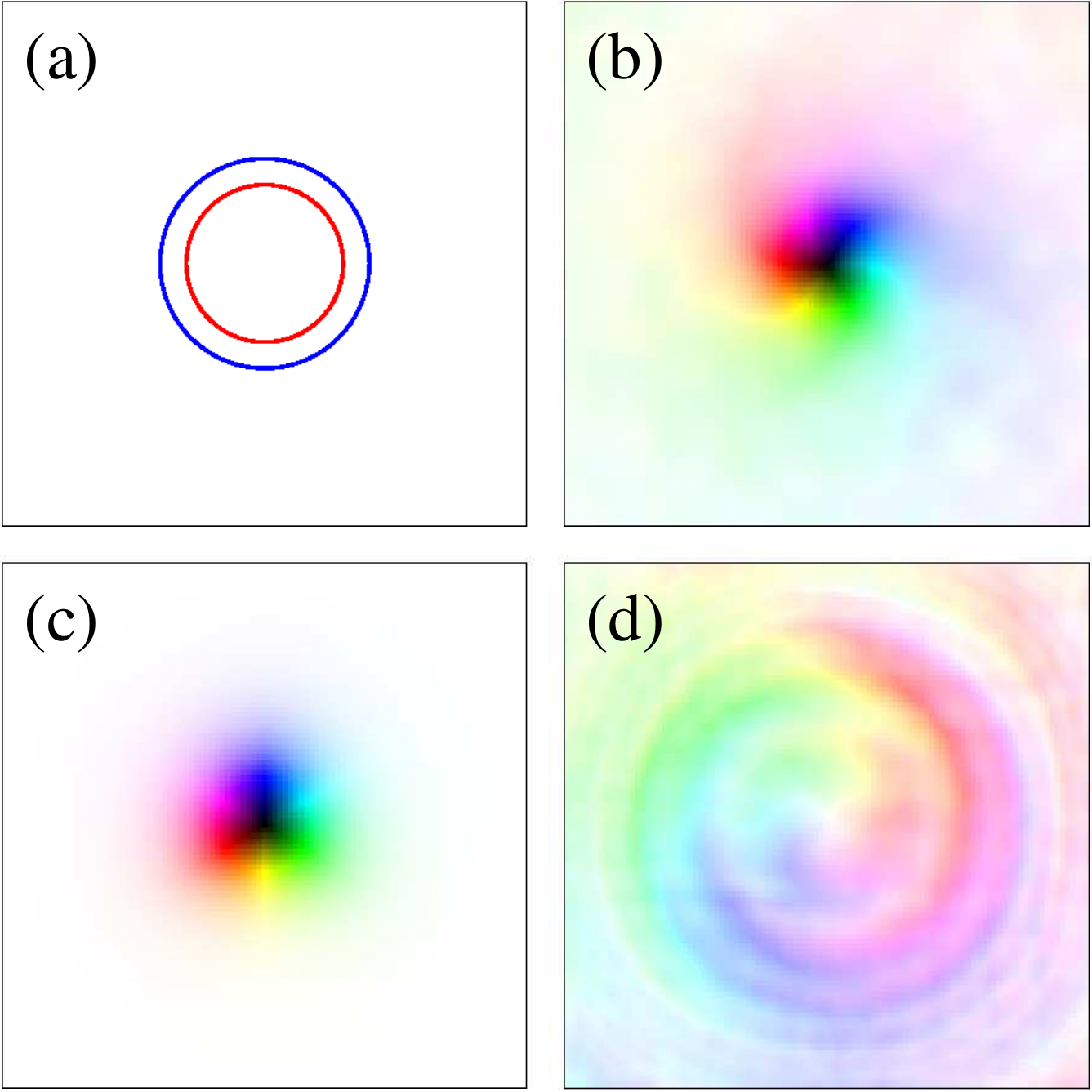}
\end{center}
\caption{(Color online) Writing and erasing an Sk by heat.  
A parameter set $\{J=1.0, D=0.15, h_0=0.02, k_BT=1.5, t_{heating}=200, \alpha=0.01, L=300\times300\}$ 
and the periodic boundary condition (PBC) are used.  
(a) Initial state without Sk. A 100$\times$100-region in the system is shown.  
Red and blue circles represent the heating spot sizes 
(the radii are 15 and 20, respectively, in this case) 
for Sk writing and erasing, respectively.  
(b) Sk writing by local heat.  
(c) Initial state with an Sk.  
(e) Sk erasing by local heat.}
\label{heat}
\end{figure}

\subsubsection{Cell memory: magnetic field control.}
\label{sec:hcontrol}

One can write and erase an Sk by changing the magnetic field strength 
as a function of time in the presence of the system boundary as shown in Fig.~\ref{h1cell} 
(see also the Movie 3 in the Supplementary Informations):  
A parameter set $\{J=1.0, D=0.18, h_0=0.03, \alpha=0.04, L=50\times50\}$ and 
open boundary condition (OBC) are used. 
A trapezoidal magnetic field pulse is applied 
inside the red rectangle $R=20\times30$ in the system 
(see Fig.~\ref{h1cell}(a)), 
so that the magnetic field  
$h_R=h_{\bm r\in R}$ is reduced (the minimum is $h_1=0.01$) 
during $0\leq t\leq3000$ (see Fig.~\ref{h1cell}(e)) and 
for $\bm r\notin R$, $h_{\bm r}=h_0$ (bias field).  
The region $R$ involves the upper edge of the system and the Sk appears from the edge 
as Figs.~\ref{h1cell}(a)$\rightarrow$(b)$\rightarrow$(c) by the magnetic field pulse.  
As seen in Fig.~\ref{h1cell}(b), in particular, the nucleation center for the Sk, the black area, 
penetrates from the upper edge 
(see also the Movie 3 in the Supplementary Informations for the whole process).    
Note that the intensity of the pulse ($h_0-h_1=0.02$) is two orders of magnitude smaller than $J$.  
It is confirmed that the magnetic field pulse with this intensity can not change 
the magnetic texture at all when $R$ is deep inside of the magnet 
and far from the edge of the system.  
At the edge of the system, the magnetic moments are lying parallel to the boundary and 
the spatial discontinuity reduces 
the topological stability, thus the Sk writing is achieved by a small magnetic field pulse.  

\begin{figure}[t]
\begin{center}
\includegraphics[scale=0.7,angle=0,clip]{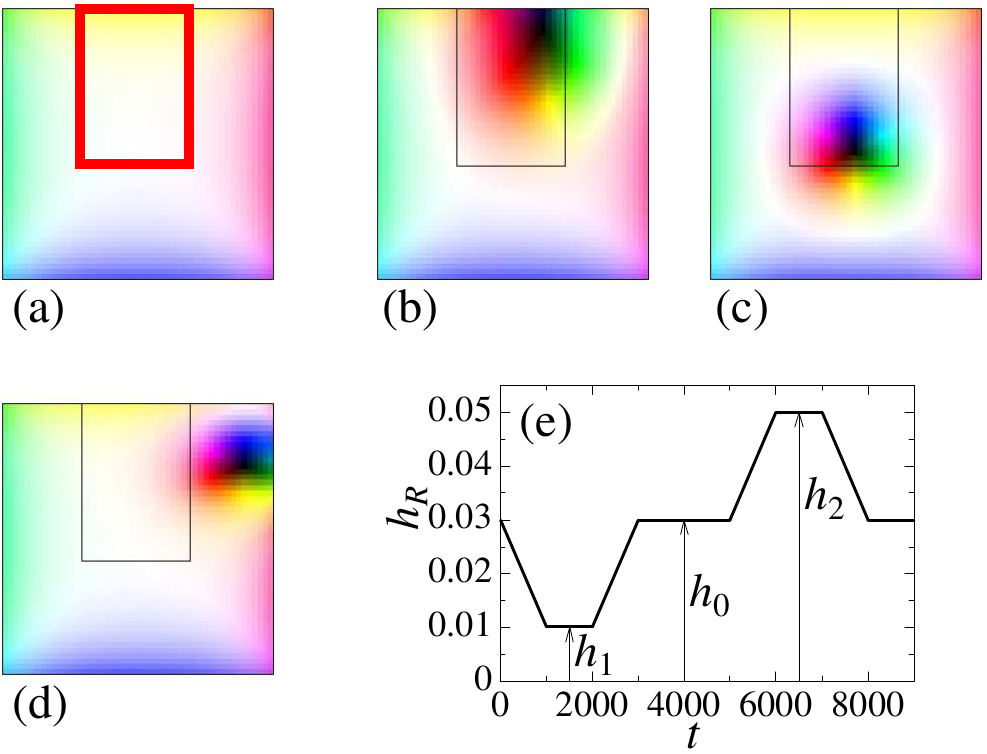}
\end{center}
\caption{(Color online) Writing and erasing an Sk by magnetic field.  
(a) Nano-sized chiral magnet without Sk. 
This is the initial state.  
The red rectangle represents the region $R=20\times30$ where the magnetic field pulse is applied. 
A parameter set $\{J=1.0, D=0.18, h_0=0.03, h_1=0.01, h_2=0.05, \alpha=0.04, L=50\times50\}$ and 
open boundary condition (OBC) are used. 
Time evolution of the magnetic texture in Sk writing is shown by the snapshots at 
(b) $t=1200$ and (c) 4000.  
Time evolution of the magnetic texture in Sk erasing is expressed by the snapshots at 
(c) $t=4000$ and (d) 7700. 
(e) Time dependence of the magnetic field $h_R$ in the region $R$.}
\label{h1cell}
\end{figure}

For the Sk erasing, in the present example,  
a positive magnetic field pulse with the maximum $h_2=0.05$ 
is applied at $5000\leq t\leq8000$ 
as shown in Fig.~\ref{h1cell}(e).  
Here, the Sk becomes unstable 
and is pushed out from the region $R$.  
As implied from Eq.~(\ref{eqn:1}), the Sk moves 
along the boundary of region $R$ due to the magnetic-field gradient. 
Consequently, the Sk turns in the dead-end 
at upper-right corner of the system as shown in Fig.~\ref{h1cell}(d),  
and eventually disappears from the system 
(see also the Movie 3 in the Supplementary Informations for the whole process).  
For the Sk erasing, the $narrow$ dead-end designed in this device structure is of 
crucial importance, i.e., the Sk does not disappear 
if the space between the region $R$ and the right edge of the system 
is as large as the Sk size.

\subsubsection{Cell memory: electric field control.}
\label{sec:econtrol}
\begin{figure}[b]
\begin{center}
\includegraphics[scale=0.7,angle=0,clip]{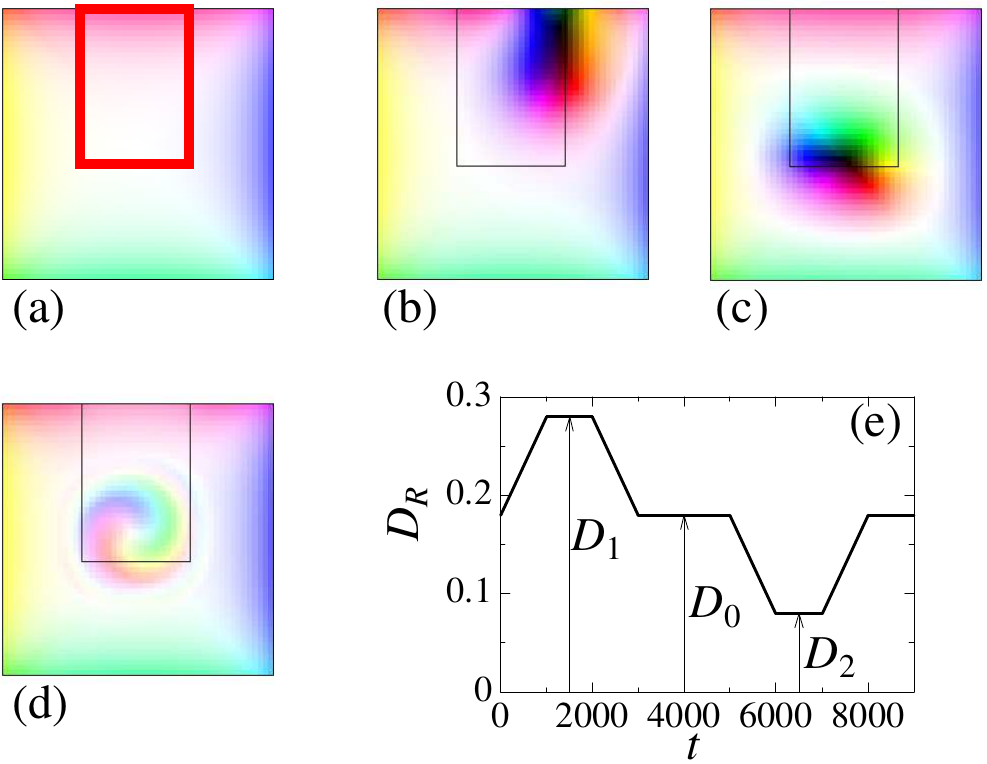}
\end{center}
\caption{(Color online) Writing and erasing an Sk by electric field.  
(a) Nano-sized chiral magnet without Sk. 
This is the initial state.  
The red rectangle represents the region $R=20\times30$ where the electric field pulse is applied. 
A parameter set $\{J=1.0, D_0=0.18, D_1=0.28, D_2=0.08, h_0=0.03, \alpha=0.04, L=50\times50\}$ 
and OBC are used. 
Time evolution of the magnetic texture in Sk writing is shown by the snapshots at 
(b) $t=1200$ and (c) 4000.  
Time evolution of the magnetic texture in Sk erasing is expressed by the snapshots at 
(c) $t=4000$ and (d) 5820. 
(e) Time dependence of the DM interaction $D_R$ in the region $R$.}
\label{e1cell}
\end{figure}

Here, we examine the Sk writing/erasing by the change in DM interaction 
with a fixed bias magnetic field.  
Figure \ref{e1cell} summarizes the result using the same numerical condition 
as in Sec.~\ref{sec:hcontrol}, 
except for the DM interaction (see also the Movie 4 in the Supplementary Informations): 
We study N\'eel Sk in this calculation, i.e., 
$\bm D_{1,\bm r}=-D_{\bm r}\bm e_y$ and $\bm D_{2,\bm r}=D_{\bm r}\bm e_x$.  
For $\bm r\notin R$, we fix $D_{\bm r}=D_0=0.18J$.  
The bias magnetic field $h_0=0.03J$ which is slightly larger 
than $h_c\approx 0.78\times(D_0^2/J)\approx0.025J$~\cite{Mochizuki,Iwasaki1}.  
The expression of $h_c$ indicates that 
the increase in DM interaction has a similar effect to the decrease in magnetic field.  
For $0\leq t\leq3000$, the DM interaction $D_R=D_{\bm r\in R}$ is 
increased as shown in Fig.~\ref{e1cell}(e).  
Upon this change in $D_R$, an Sk is created as Fig.~\ref{e1cell}(a)$\rightarrow$(b)$\rightarrow$(c) 
and this behavior is similar to the Sk creation in Sec.~\ref{sec:hcontrol}.  

On the other hand, the Sk annihilation occurs without the drift motion 
and the behavior differs from the case of magnetic-field change in Sec.~\ref{sec:hcontrol}.  
For $5000\leq t\leq7000$, the $D_R$ is decreased as shown in Fig.~\ref{e1cell}(e).
This decrease in $D_R$ with fixed $h_0$ leads to instability of the Sk and 
the size of Sk is proportional to $D_R/J$.  
Therefore Sk shrinks in size and finally disappears.  
It is noted that the position of Sk does not change 
in the erasing process in sharp contrast to the case of magnetic field change. 
This is due to the fact that the electric field gradient does not produce 
the force acting on the Sk since the energy of the Sk does not depend on $D$~\cite{foot1}.  

\subsubsection{Cell memory: Current control.}

\begin{figure}[b]
\begin{center}
\includegraphics[scale=0.6,angle=0,clip]{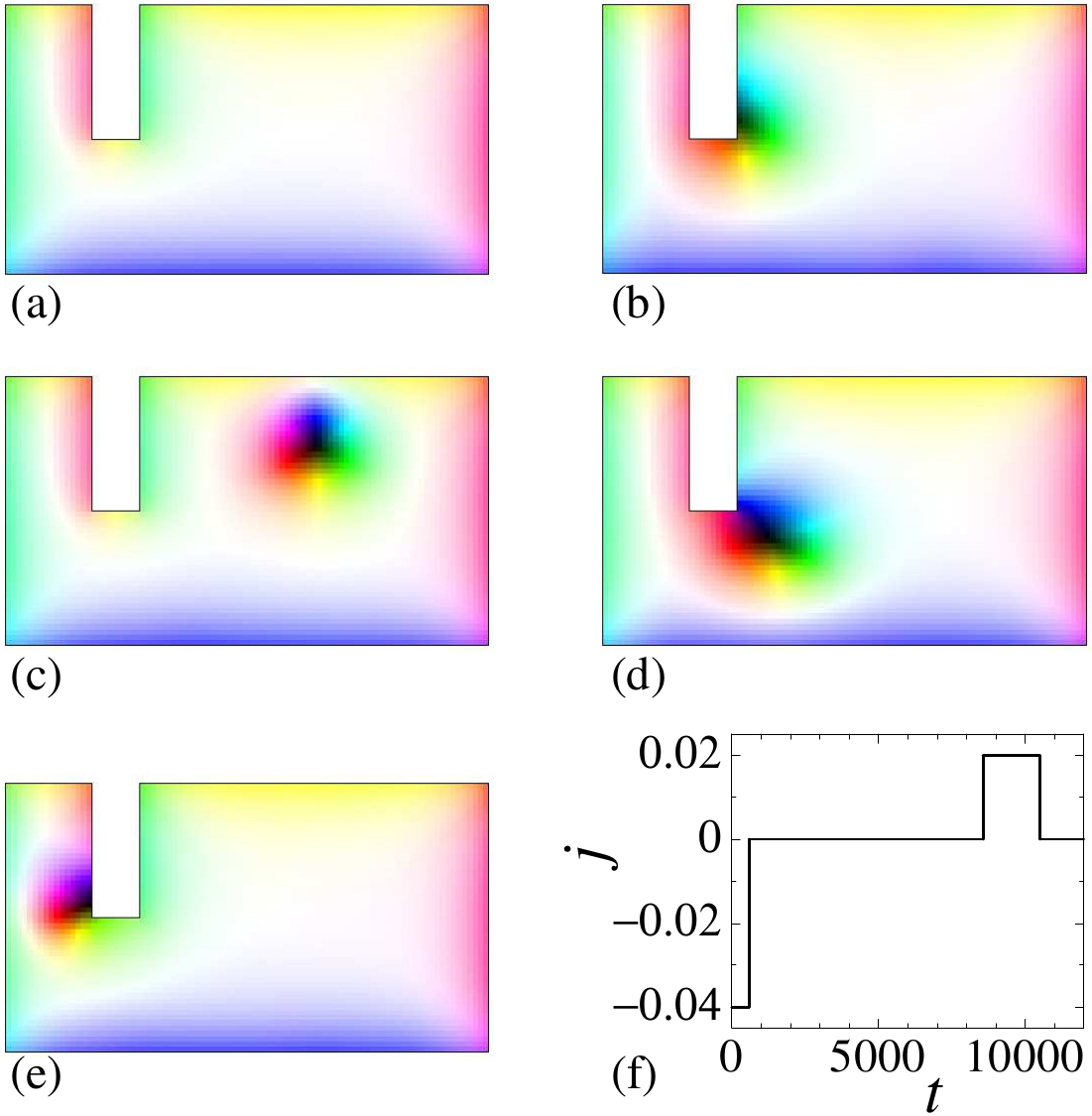}
\end{center} 
\caption{(Color online) Writing and erasing an Sk by electric current.  
A parameter set $\{J=1.0, D=0.18, h_0=0.0278, k_BT=0.0, \alpha=\beta=0.04\}$ and  
OBC are used.  
The $90\times50$ system has a notch $10\times50$ 
whose left edge is at 16 from the left-end of the system.  
(a) The system without Sk. 
This is the initial state.  
Time evolution in Sk writing is shown by the snapshots at 
(b) $t=300$ and (c) 1500.  
Time evolution in Sk erasing is shown by the snapshots at 
(d) $t=10350$ and (e) 10800.  
(f) Time dependence of the electric current density at the right end of the system.  
See ref.~\citen{Iwasaki2} for similar calculations.
}
\label{j1cell}
\end{figure}

Here, we examine the Sk writing/erasing by an electric current.  
A previous study~\cite{Iwasaki2} shows that  
the notch structure in the constricted geometry 
is effective to create an Sk.  
This geometry is also important 
to annihilate Sk by lower electric current density.  
Figure \ref{j1cell} shows the result 
with a parameter set $\{J=1.0, D=0.18, h_0=0.0278, k_BT=0.0, \alpha=\beta=0.04\}$ and OBC 
(see also the Movie 5 in the Supplementary Informations).  
As shown in Fig.~\ref{j1cell}(a), in the $90\times50$ system, we design a notch structure $10\times50$ 
whose left edge is at 16 from the left-end of the system.  
In the system with a notch structure, 
the spatial dependent $\bm j_{\bm r}$ 
is determined by the current conservation law div$\bm j_{\bm r}=0$~\cite{Iwasaki2}, 
and $j$ represents the magnitude of the electric current density at the right end of the system 
($j >0$ for the current flows from left to right).  
For $0\leq t\leq600$, an electric current pulse  
with a maximum $|j|=0.04$ (see Fig.~\ref{j1cell}(d)) is applied and an Sk is created.  
The time evolution is shown in Figs.~\ref{j1cell}(a)$\rightarrow$(b)$\rightarrow$
(c).  
Here the edge of the system plays an essential role again: 
Due to the DM interaction, a substantial in-plane component of magnetic moment appears 
along the edge of chiral magnet and gives rise to a magnetic texture being compatible with Sk  
at the corner of the notch. From the specific corner, an Sk is created 
with a help of electric-current induced STT.  
Note that the consistency between magnetic texture along 
the corner and winding of Sk is essential to the creation.  
At this specific corner, the Sk is also easier to be destroyed 
than other part of edge of the system.  
The Sk erasing is caused by an electric current pulse opposite to the case of Sk writing.  
In the geometry of Fig.~\ref{j1cell}, 
$j=0.02$ with a time-interval $8550\leq t\leq10500$, 
is applied:    
First, as seen in Fig.~\ref{j1cell}(d), 
the approaching Sk to the notch structure is trapped at the corner 
where the Sk appears in the writing process (see Fig.~\ref{j1cell}(b)).   
Later on, the disappearing Sk runs along the edge of the notch and 
turns in the designed narrow dead end at the upper-left corner of the system 
as shown in Fig.~\ref{j1cell}(e).  
Finally, the Sk disappears from the system 
(see also the Movie 5 in the Supplementary Informations for the whole process).  
The shape of notch structure is of crucial importance to design the stability of Sk, i.e., 
the topological surgery by the electric current controlled Sk memory.

\subsection{Drive methods}
\label{sec:drive}
The STT by (spin-polarized) electric current drives 
Sk motion~\cite{Iwasaki1,Iwasaki2,Jonietz,YuXZ12,Sampaio,Iwasaki_gliding,Tomasello} 
and the current is much smaller 
than that for the motion of ferromagnetic domain walls 
(DWs)~\cite{Slonczewski,Berger,Maekawa,Parkin,Ono,Yamaguchi,KJKim}.  
This advantage is, however, diminished by the confining geometry; 
in the $narrow$ Sk lead-track, the magnitude of required electric current for Sk motion 
becomes as large as that in the case of DWs 
when the electric current $\bm j$ is applied parallel to the lead-track.  
This is due to the effect of the Sk confining potential 
(expressed by $\nabla U$ in Eq.~(\ref{eqn:1})) 
at the edge.  
In the confining geometry, on the other hand, 
an efficient Sk driving force emerges at the edge, i.e.,  
the Sk $gliding$-motion along the edge  
by electric current in $perpendicular$ direction~\cite{Sampaio,Iwasaki_gliding}.  
Figure \ref{edge} manifestly demonstrates the efficiency of the Sk gliding-motion 
(see also the Movie 6 in the Supplementary Informations):  
Here, the parameter set is $\{J=1.0, D=0.18, h_0=0.0278, L=600\times50\}$.  
In horizontal and vertical directions, PBC and OBC are used, respectively. 
Figure \ref{edge}(a) is the initial state where an Sk is in a narrow lead-track.  
The Sk moves by a uniform electric current density 
$j=|\bm j_{\bm r}|=0.001$ with a duration $t=30000$ 
in parallel ((b) and (c)) and perpendicular ((d) and (e)) directions, respectively.    
In the case of parallel current, the Sk begins to move by the STT but 
the Sk confining potential in this narrow lead-track immediately suppresses the Sk motion, and   
eventually the Sk velocity becomes proportional to $\sim\beta/\alpha$.  
Therefore, the Sk moving-distance for $\beta/\alpha=1$ shown in Fig.~\ref{edge}(b) is larger 
than that for $\beta/\alpha=0$ shown in Fig.~\ref{edge}(c). 
In the case of perpendicular current, however, $\beta$ is totally irrelevant.    
Figures \ref{edge}(d) and (e) shows the results of the Sk gliding motion 
for $\beta/\alpha=1$ and 0, respectively.   
Note that the Sk moving-distance in the gliding Sk motion is much larger than that in Fig.~\ref{edge}(b).  
The velocity of the gliding Sk motion is enhanced by a factor $\sim$$1/\alpha$ 
in comparison to that in parallel current case.   
In addition to the large STT in the gliding Sk motion, 
the perpendicular condition gives a significant flexibility for the Sk-device design discussed below.   
\begin{figure}[t]
\begin{center}
\includegraphics[scale=0.55,angle=0,clip]{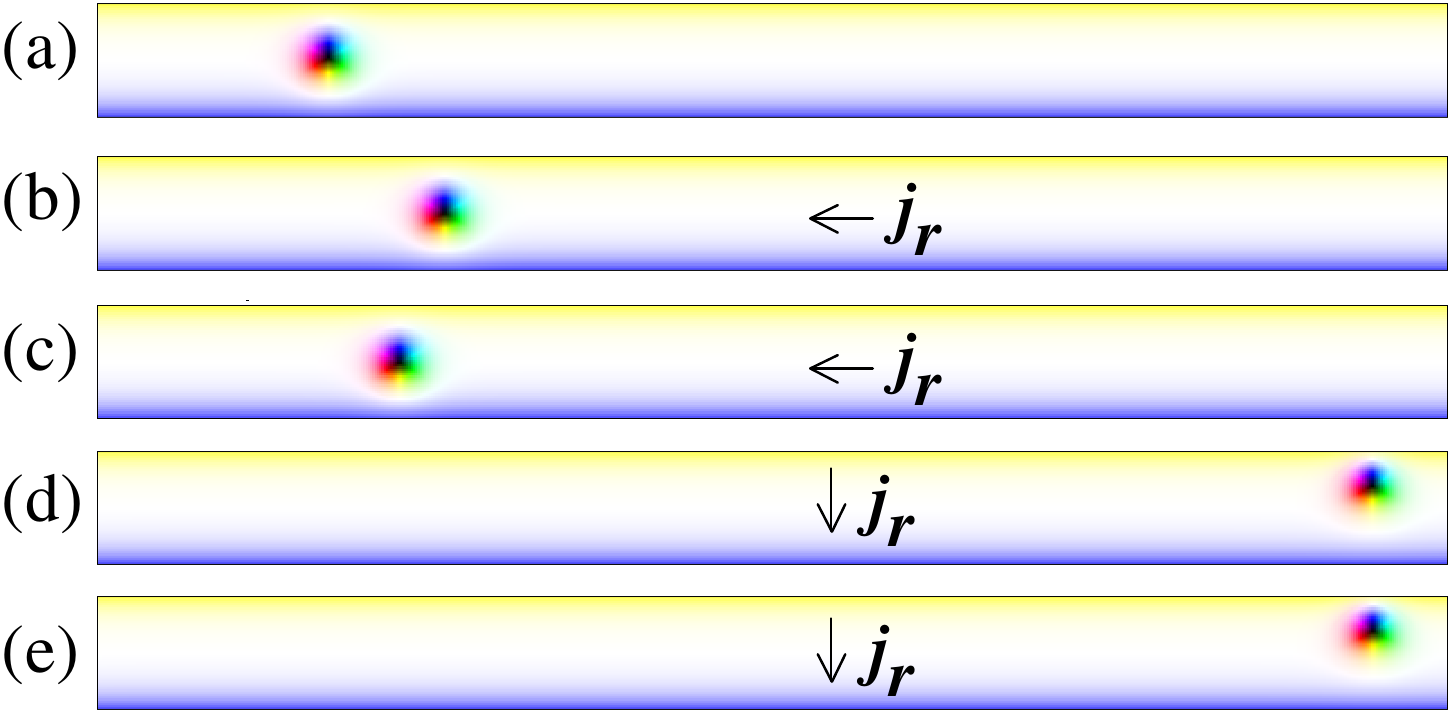}
\end{center}
\caption{(Color online) Parallel and edge gliding motions of Sk by STT. 
A parameter set $\{J=1.0, D=0.18, h_0=0.0278, L=600\times50\}$ is used.
In horizontal and vertical directions, PBC and OBC are applied, respectively.     
(a) Initial state.  
(b) The moved Sk at $t=30000$ in parallel configuration  
with a uniform electric current density $j=|\bm j_{\bm r}|=0.001$ and $\alpha=\beta=0.04$.  
(c) The same as (b) but $\{\alpha=0.04, \beta=0.0\}$.  
(d) and (e) The edge gliding motion: 
The moved Sk at $t=30000$ in perpendicular configuration with $j=0.001$ for 
(d) $\alpha=\beta=0.04$ and (e) $\{\alpha=0.04, \beta=0.0\}$.  
See ref.~\citen{Iwasaki_gliding} for similar calculations.  
}
\label{edge}
\end{figure}

\subsection{Read methods}
Now the methods to read out the presence or absence of the Sks are discussed. 
One way is to use the tunnelling magnetoresistance (TMR). 
Consider the tunnel junction composed of the ferromagnetic (F) and the skyrmionic thin films.   
The tunnelling conductance $G$ of the junction is given by the expression 
\begin{equation}
G \propto \sum_{n, \bm{k}} |<\!\bm{k} \sigma | H | n\!> |^2  \delta(E_n-E_F) 
\delta(\varepsilon_{k \sigma}-E_F) 
\label{eq:G}
\end{equation}
where $E_F$ is the Fermi energy,  $|n\!>$ and $E_n$ are the eigenstate 
and its energy eigenvalue of the skyrmionic system, 
while $\varepsilon_{\bm{k} \sigma}$ and $|\bm{k} \sigma\!>$ are those for the ferromagnet.  
The spin configuration in skyrmionic system affects $|n\!>$ and hence $G$. 
Suppose the F-system is perfectly spin polarized perpendicular to the film, i.e., 
$\sigma=+1$ in Eq.~(\ref{eq:G}).  
In the absence of Sks, the skyrmionic system is also perfectly spin polarized, 
and $G$ has the maximum value $G_0$ in this case. 
When $N$ Sks are introduced into skyrmionic system, the fraction 
$\cong N \pi \xi^2/ A$ ($\xi$: Sk size, $A$: sectional area of the F and skyrmionic thin-film-junction) 
of the spins are reversed to down direction, 
which cannot contribute to the conductance 
since it has no overlap integral with the wavefunction in F-system. 
Therefore, the reduction of the conductance $\Delta G$ is estimated as
\begin{equation}
\frac{\Delta G}{G_0} \cong N \pi \xi^2/ A.
\end{equation}
When the area $A$ is comparable to the size of one Sk, 
this ratio can be a reasonable fraction of unity, 
and hence can be a sensitive probe of the presence or absence of the Sks. 

The other method for the detection is to utilize 
the topological Hall effect (THE)~\cite{Neubauer,Schulz,Iwasaki3}. 
As described above, 
the Sk acts as the effective magnetic flux on the conduction electrons coupled to it. 
Each Sk is associated with the unit flux $\phi_0 = 2\pi\hbar/e$ 
in the strong limit of spin-charge coupling~\cite{reviewNT}. 
This effective magnetic field scatters the incoming wave asymmetrically 
between the right and left directions. 
When the wavenumber $k$ satisfies the condition $k \xi \cong \pi/2$, 
the skew scattering angle has a peak value of the order of 60 degree 
and decreases as $\sim 1/(k\xi)$ for $k \xi >>1$~\cite{Iwasaki3}. 
This skew scattering produces the Hall voltage perpendicular to the direction of the current. 
When the periodic array of the Sks, i.e., SkX, is formed, 
an almost uniform effective magnetic field $<\!b_z\!>$ is present as  
\begin{equation}
<\!b_z\!> = \frac{\sqrt{3} \phi_0}{2 \lambda^2} 
\end{equation}
where $a_s = 2 \lambda/\sqrt{3}$ is the lattice constant of the triangular SkX. 
This produces THE~\cite{Neubauer,Schulz,Iwasaki3} replacing 
the external magnetic field by this effective magnetic field.  

\section{Some models for Sk memory}
\label{Sec:somemodels}

\subsection{Cell memory}

In Sec.~\ref{write/erase}, we discussed the Sk writing/erasing  
with the simulation results  
Figs.~\ref{heat}, \ref{h1cell}, \ref{e1cell} and \ref{j1cell}.  
These results also provide a set of design principles 
for the Sk cell memories; the size of the system and 
local area for external forces are scaled by the Sk size $\xi$, 
the intensity of the external forces are 
scaled by $J$ and operation time is scaled by $1/(\gamma J)$.  
If we assume $J \sim 10^{-3}$ eV, $1/(\gamma J) \sim 0.7$ ps. 
Therefore, the Sk writing/erasing time $\Delta t$ is 
of the order of nano or pico seconds. 
For example, in the electric current control device discussed in Fig.~\ref{j1cell}, 
the Sk writing energy is expressed to be $\Delta T\times\rho j^2\times Ad$ 
where $\rho$ is electrical resistivity, $A$ and $d$ are the area and thickness of the film sample.  
If we assume the parameter set~\cite{Iwasaki1,Iwasaki2,Jonietz,YuXZ12}, 
$\{\Delta t=200$ps, $\rho=100\mu\Omega$cm, $j=1\times10^{7}$A/cm$^2$, 
$A=(La^2=)45$nm$\times25$nm, $d=10$nm$\}$, 
$E_{cost}$=0.0225fJ is estimated for an Sk writing.  
The F exchange coupling $J$ should be increased for the room temperature Sk-hosting magnet.  
The increase in $J$ causes an increase in $E_{cost}\propto J$ 
(assuming the same $\rho$) 
because $\Delta t\propto J^{-1}$ and $j\propto J$.  Nevertheless,  
$E_{cost}$ is still several orders of magnitude smaller than 0.09pJ which is estimated 
for writing electric power in the best condition of the currently developing 
STT-MRAM~\cite{Kitagawa}.  
Moreover, the Sk memory designed in the Sec.~\ref{sec:econtrol} works by 
an external electric field, giving rise to a principle for a further 
low-energy-cost Sk memory device.

\vskip1cm
\subsection{Slide-switch memory}

\begin{figure}[b]
\begin{center}
\includegraphics[scale=0.7,angle=0,clip]{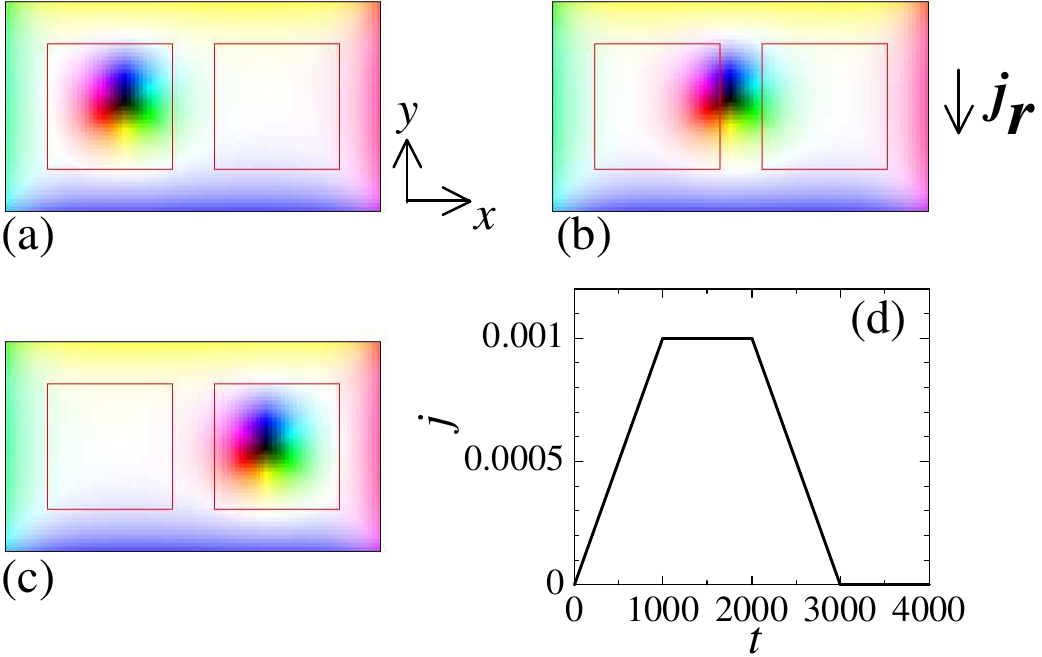}
\end{center}
\caption{(Color online) Sliding-switch memory. 
The two red rectangles, $30\times30$ each in the system with the size $L=90\times50$, 
specifies the spatial dependence in static (bias) magnetic field $h_{\bm r}$, 
i.e., $h_{\bm r}=h_R$ inside and $h_{\bm r}=h_0$ outside. 
A parameter set $\{J=1.0, D=0.18, h_0=0.03, h_R=0.029, \alpha=0.04, \beta=0.0\}$ and OBC are used. 
(a) Initial state.   
(b) The result of gliding Sk motion at $t=3000$ by an electric current pulse 
with the maximum current density $j=|\bm j_{\bm r}|=0.001$.  
(c) The magnetic texture at $t=8000$.  
(d) Time dependence of $j$.  For $t>3000$, $j=0$.  
}
\label{2cell}
\end{figure}

A possible device design of the Sk memory without Sk writing/erasing is shown in Fig.~\ref{2cell}, 
where the red rectangles represent the stable areas for the Sk and 
the stay of the Sk in left or right red rectangle corresponds to `1' or `0' states 
(see also the Movie 7 in the Supplementary Informations).   
A parameter set $\{J=1.0, D=0.18, h_0=0.03, \alpha=0.04, \beta=0.0, L=90\times50\}$ and OBC are used. 
The local minima of the potential for the Sk is prepared by the spatial dependence of $h_{\bm r}$, i.e.,   
the static (bias) magnetic field inside the red rectangles ($30\times30$ each) $h_R=0.029$ is weaker 
than that in other region $h_0=0.03$. 
For the switching between `1' to `0', the current driven motion of Sk, in particular 
the gliding Sk motion discussed in Sec.~\ref{sec:drive}, is efficient.   
In the case shown in Fig.~\ref{2cell}, the electric current pulse in $y$ direction 
with the maximum current density $j=|\bm j_{\bm r}|=0.001$ is enough to switch the state,   
whereas this intensity is too small to do it in the case that $\bm j_{\bm r}\parallel\bm e_x$.  
Note that $j=0$ for $3000< t$, i.e., the electric current pulse is turned off before the Sk 
goes into the right stable area 
and inertia~\cite{Iwasaki_gliding} in the Sk gliding-motion finalizes the switching 
(see also the Movie 7 in the Supplementary Informations for the whole process).  

\subsection{Racing-circuit memory}

\begin{figure}[t]
\begin{center}
\includegraphics[scale=0.45,angle=0,clip]{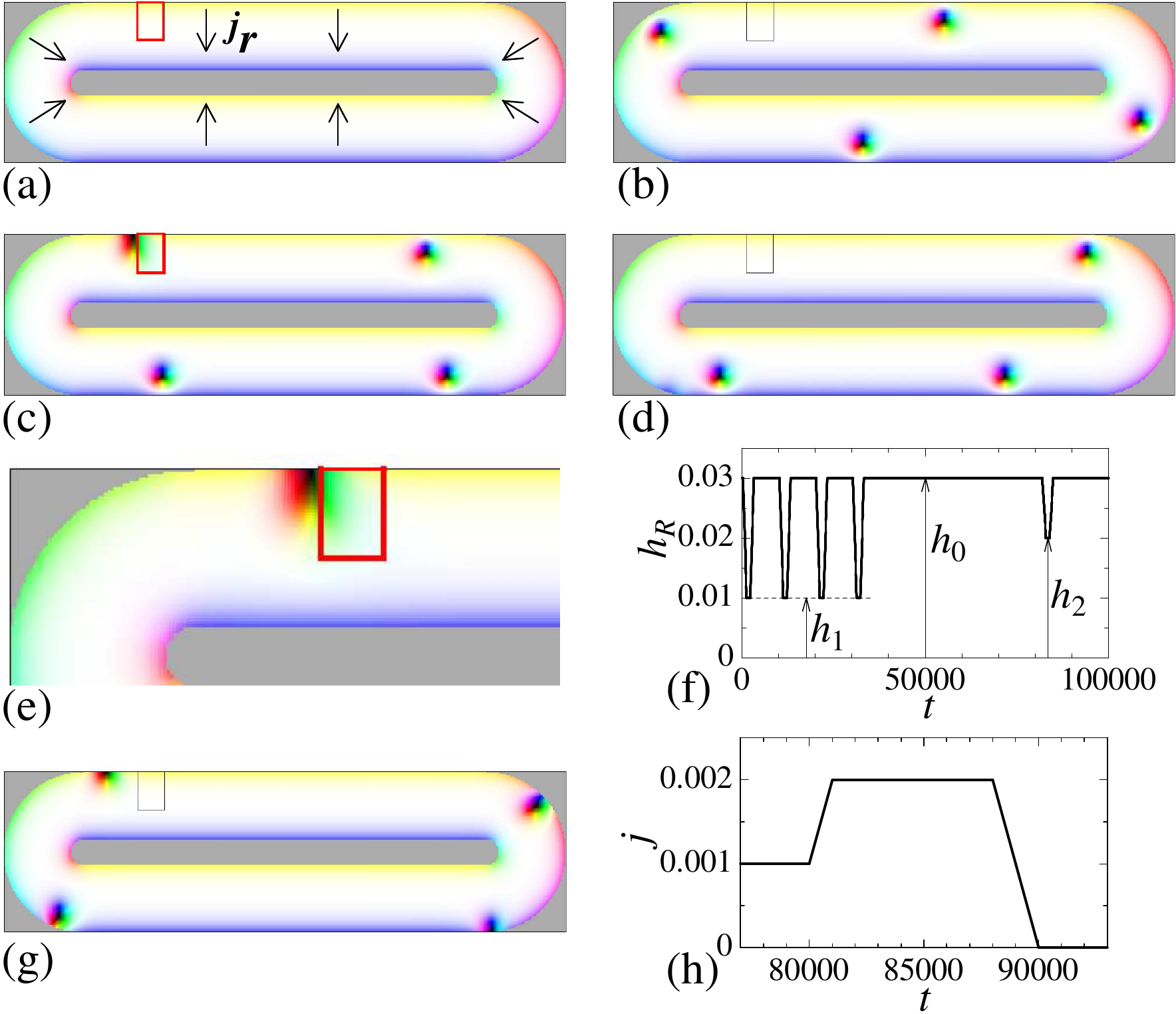}
\end{center}
\caption{(Color online) Racing-circuit memory. 
A parameter set $\{J=1.0, D=0.18, h_0=0.03, h_1=0.01, h_2=0.02, \alpha=0.04, \beta=0.0\}$ and OBC are used. 
In the $420\times120$-rectangle-area, the chiral magnet with width of $50$ forms a closed lead-track for 
the moving Sks.  
(a) Initial state without Sk.  
Electric current $\bm j_{\bm r}$ with a magnitude $j=|\bm j_{\bm r}|=0.001$ 
is flowing from outside to inside of the lead-track. 
In the red rectangle ($20\times30$), writing/erasing magnetic field pulse $h_R(t)$ is applied. 
(b) Four Sks created by magnetic field pulses are moving 
in clockwise direction along outer-edge of the circuit. 
For the writing of an Sk, a magnetic field pulse as in Sec.~\ref{sec:hcontrol} 
i.e., a trapezoidal magnetic field pulse with 
an intensity $h_0-h_1=0.02$ at a duration $\Delta t=3000$ is used.  
Time evolution of the Sk erasing by an magnetic field pulse 
is shown by the snapshots at (c) $t=83200$ and (d) 85000.  
The difference between the Sk writing and erasing magnetic field pulse is 
the intensity, i.e., 
an intensity $h_0-h_2=0.01$ at $81900\leq t\leq84900$ is used for the Sk erasing.  
(e) A close-up of (c).  
Time dependence of $h_R$.   
(g) Reset procedure: snapshot at $t=88300$.  
(h) Time dependence of $j$ in reset procedure.  
}
\label{racing}
\end{figure}

The gliding Sk motion also enables the flexible device design; 
the $closed$ circuit can be the lead-track for the Sk motion which would be impossible 
with the parallel current direction discussed in Sec.~\ref{sec:drive}.  
Figure \ref{racing} shows an example where the chiral magnet 
with the width of $50$ forms a closed lead-track for 
the moving Sks in the $420\times120$-rectangle-area.  
A parameter set $\{J=1.0, D=0.18, h_0=0.03, \alpha=0.04, \beta=0.0\}$ and OBC are used. 
An electric current $\bm j_{\bm r}$ with a magnitude $j=|\bm j_{\bm r}|=0.001$ 
is flowing from outside to inside of the Sk lead-track and causes the motion of the Sk 
along the circuit wall.  
Here the magnetic field control is used for the Sk creation and annihilation 
(see also the Movie 8 in the Supplementary Informations),   
while all the writing/erasing methods of the Sk discussed in Sec.~\ref{write/erase} are available.  
In the red rectangle region ($20\times30$) shown in Fig.~\ref{racing}(a), 
the magnetic field pulse $h_R(t)$ is applied.  
For an Sk-writing magnetic field pulse,  
the condition in Fig.~\ref{h1cell}, i.e., a trapezoidal magnetic field pulse with 
an intensity $h_0-h_1=0.02$ at a duration $\Delta t=3000$ is employed.  
By the four pulses at $0\leq t\leq 33000$, 
four Sks are created and keep moving in clockwise direction 
(see Figs.~\ref{racing}(b) and (f)).   
For the erasing of Sk, a $weak$ magnetic field pulse, 
which is applied at $81900\leq t\leq84900$ and 
whose intensity $h_0-h_2=0.01$ is half of the Sk writing pulse in this case, 
is used.  
Figures \ref{racing}(c)$\rightarrow$(d) show the snapshots of the Sk erasing 
by the weak magnetic field pulse and Fig.~\ref{racing}(e) 
is a closeup of Fig.~\ref{racing}(c), where the Sk disappears through the upper edge 
(see the Movie 8 in the Supplementary Informations for the whole process).  
The intensity of the erasing pulses is not strong enough to create the Sk, however, 
the local magnetic field reduction brings about an attractive force to the Sk.  
Because the Sk motion follows the Eq.~(\ref{eqn:1}), the Sk moves perpendicular to the attractive force.  
Therefore, the Sk approaching to the red rectangle Sk jumps out of the system through the upper edge 
by the weak magnetic field pulse, as shown  
in Figs.~\ref{racing}(c) and (e).  

The confining potential for the Sk at the edge is finite, and hence 
there exists a critical value in the electric current density $j=|\bm j|$ for the gliding Sk motion.  
This is useful for $reset$ procedure.  
Figures \ref{racing}(g) and (h) show a snapshot of the disappearing Sks 
and the time-dependence of the electric current pulse for the reset procedure, 
respectively (see also the Movie 9 in the Supplementary Informations):  
For $0\leq t\leq80000$, the same numerical condition as that in the previous case, Figs.~\ref{racing}(a-f),
is used.   
All the Sks are pushed out by the electric current pulse at $80000\leq t\leq9000$ 
with a maximum $j=0.002$ and disappears through the outer edge of the system.  
 
\section{Summary and perspectives}
\label{Sec:summary}
Here it is worthwhile to compare the characteristics 
of skyrmion memory devices with other magnetic ones 
such as bubble~\cite{MaloSlo,Chikazumi} and racetrack~\cite{Parkin} memories that also 
employ memory bits or domains in magnetic thin films. 
Table~\ref{t1} summarizes the comparison of these three magnetic memories. 
The bubble memory~\cite{MaloSlo,Chikazumi} had been developed since 1960's and were 
at market around 1980. 
The bubbles were manipulated in insulating iron garnet films 
by means of an external magnetic field and were detected 
by the magnetic field coming out of the bit by magnetoresistance or 
Hall devices. Compared with bubbles, the size of magnetic bits for 
racetrack and Sk memories are quite small that are necessary 
condition for the pursuit of high density memory. 
It should be noted that some of the knowledge obtained 
in the research of magnetic bubble device~\cite{MaloSlo,Chikazumi} 
are referred in the course of skyrmion memory research.

\begin{table}[t]
\caption{Comparison of memory devices.}
\label{t1}
\begin{tabular}{lp{4cm}p{4cm}p{4cm}l}
\Hline
 & Bubble & Race-track & Sk \\
\Hline
Size & $\sim \mu$m & $10\sim100$nm & $3\sim 100$nm\\ \hline
Write/Erase & magnetic field & Current (Spin transfer trque) 
& Current, Heat, Electric field\\
Read & Magneto-resistive sensor, Hall sensor & Magneto-tunneling junction (MTJ) 
& Topological Hall effect, MTJ, Hall sensor\\
Drive & Magnetic field & Current: $10^7$ A/cm$^2$ & Current: $10^2$ A/cm$^2$ \\
\Hline
\end{tabular}
\end{table}

When the latter two are compared, 
a most distinct feature of skyrmion device 
is the very low current density needed to drive magnetic bits, 
orders of magnitude lower than that required 
for domain wall motion in racetrack memory~\cite{Parkin}. 
This provides high potential for low energy-cost devices. 
The storage capacity of memory device is 
mostly limited by the size of the magnetic bits. 
Skyrmion memory having, comparable to or smaller size than racetrack memory, 
has a big advantage in this regard as well. 
Another important characteristic is the variety of 
tools for writing/erasing and reading the magnetic bits. 
Racetrack memory solely relies on sophisticated 
magnetic tunnel junction (MTJ). 
Spin polarized current injection through 
a tunnel barrier provides 
spin transfer torque to reverse the magnetization. 
This could also be employed for skyrmion device as discussed 
by Fert {\it et al.}~\cite{Fert}. 
However, this method usually consumes rather large energy 
because high density spin polarized current 
has to be fed through a high resistance MTJ. 
For skyrmions, as discussed in Section \ref{Sec:section2}, 
other techniques such as current flow in low resistance channel, 
local heat application, and application of electric field to control DM interaction, 
can be used to explore novel possibility for low energy-cost memory devices. 
As for reading the bit, MTJ is an excellent method and will be a choice for 
skyrmion device as well. However, other choices such as topological 
Hall effect may explore new methods to realize high performance memory device.

Since magnetic structures are used, all the devices are expected 
to have high enough endurance without any fatigue or 
imprint that are representative limiting problem for ferroelectric 
memory where atomic motion takes place. 
Due to topological protection of skyrmion, 
the retention is also expected to be as excellent as other magnetic devices. 
Therefore, skyrmion has significant potential for memory devices. 

The issues discussed in this paper have been theoretically verified 
to raise numbers of challenges to experimentalists. 
Materials choice~\cite{reviewNT} and micro-fabrication~\cite{Kubota,Kanazawa} have been investigated 
so that we can examine basic characteristics of elementary 
device actions such as write/erase, read, and drive. 
As have been the case for former challenges in semiconductor 
devices and magnetic storage, 
interplay between theory and experiments as well as basic device physics 
and application oriented developments will be the key for exploring novel 
science/technology based on skyrmions.

\acknowledgment
We are grateful to Y. Ogimoto and A. Toriumi for stimulating discussions.  
We also thank M. Ishida for technical assistances.
This work was supported by a Grant-in-Aid for Scientific Research
(No. 24360036, No. 24224009)
from the Ministry of Education, Culture, Sports, Science and
Technology (MEXT) of Japan and 
by the Funding Program
for World-Leading Innovative R\&D on Science and Technology (FIRST Program).
J.I. was supported by Grant-in-Aids for JSPS Fellows (No. 2610547).


\begin{thebibliography}{9}
\bibitem{foot1}
For the long $\xi\sim(J/D)a$ (Sk size), the stabilization energy density $\varepsilon_{Sk}$ 
due to the DM interaction is expressed to be $\varepsilon_{Sk}\sim (J/\xi^2)$.  
Therefore the stabilization energy of an Sk is $E_{sk}=\varepsilon_{Sk}\times(\pi\xi^2)\sim\pi J$ 
and is $not$ dependent on $D$.
\end{thebibliography}
\end{document}